\documentclass{PoS}
\usepackage{amsmath}
\usepackage{wrapfig}
\usepackage{amssymb}
\usepackage{lineno}
\usepackage{color}
\usepackage{graphicx}
\usepackage{relsize}
\def\babar {{{\mbox{\slshape B\kern-0.1em{\smaller A}\kern-0.1em B\kern-0.1em{\smaller A\kern-0.2em R }}}}}

\title{Charmonium-like States at  \babar}

\ShortTitle{Charmonium-like States at  \babar}

\author{\speaker{Valentina Santoro}\\
        INFN Ferrara, via Saragat~1, 44122 Ferrara, Italy\\
        E-mail: \email{vsantoro@slac.stanford.edu}}

\abstract{We present recent results on charmonium and charmonium-like states from the BaBar B-factory located at the PEP-II asymmetric energy $e^{+}e^{-}$ storage ring at the SLAC National Accelerator Laboratory. }

\FullConference{36th International Conference on High Energy Physics,\\
		July 4-11, 2012\\
		Melbourne, Australia}

\begin{document}

\section{Introduction}
The charmonium spectrum consists of eight narrow states below the open charm threshold (3.73~GeV) and several tens of states above that. 
Below the threshold almost all states are well established, on the other hand very little is known at higher energy. In addition to that there have been discoveries~\cite{qr} of several new charmonium-like states above the threshold. While some of them appear to be consistent with conventional charmonium others don't seem to behave like standard meson and could be made of a larger number of constituent quarks.\\
In the next sections we will review recent \babar results on this field.

\section{Study of the $J/\psi \omega$ in two-photon interactions}
The Y(3940) was observed for the first time by Belle \cite{Abe:2004zs} in B decays
 and then confirmed by \babar  \cite{babary3940)}. In a re-analysis  \cite{X3872} of the \babar data sample the precision of the Y(3940) parameters was improved and was also found evidence for the decay $X(3872) \to J/\psi \omega$. This confirmed an earlier Belle claim  \cite{X3872belle}
 for the existence of this decay mode. A subsequent Belle paper \cite{X3915belle} reports the evidence of a structure in the process $\gamma \gamma \to J/\psi \omega$ that they dubbed the  X(3915) with mass and width values similar to those obtained for the Y(3940) by \babar \cite{babary3940)}. \babar has recently performed a study of the process  $\gamma \gamma \to J/\psi \omega$~\cite{X3915BaBar}
 to search for the X(3915) and the X(3872) using a data sample of 519 $\mathrm{fb^{-1}}$. Figure 1 shows the reconstructed $J/\psi \omega$ mass distribution after all the selection criteria have been applied. A large peak at near $3915~{\mathrm{MeV/c^{2}}}$ is observed with a significance of 7.6 $\sigma$. The measured resonance's parameters obtained from a maximum likelihood fit are $m_{X(3915)}=(3919.4 \pm 2.2  \pm 1.6) {\mathrm MeV/c^{2}}$, $\Gamma_{X(3915)}=(13\pm 6\pm 3)$MeV. The measured value of the two-photon width times the branching fraction, is $\Gamma_{\gamma \gamma}(X(3915)) \times {\cal B}(X(3915) \to J/\psi \omega) =52\pm 10 \pm 3$~eV and $10.5 \pm 1.9 \pm 0.6$~eV for the spin hypothesis J=0 and J=2, respectively, where the first error is statistical and the second is systematic. In addition a Bayesian upper limit (UL) at 90 \% confidence level (CL) is obtained for the X(3872): $\Gamma_{\gamma \gamma}(X(3872)) \times {\cal B}(X(3872) \to J/\psi \omega) <1.7 $~eV, assuming J=2.

\begin{figure}[htb]
  \centering
  \includegraphics[height=.22\textheight]{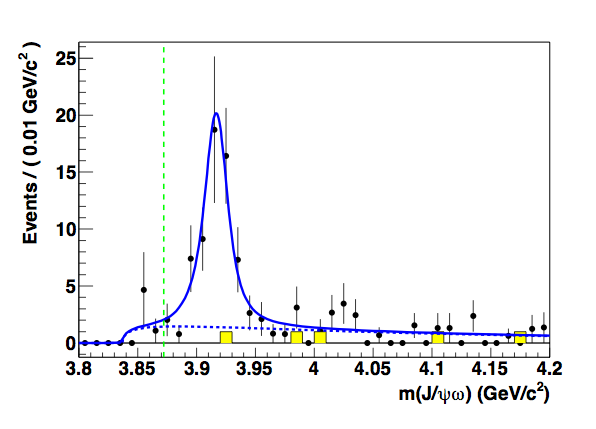}
  \label{fig1}
  \caption{The efficiency-corrected invariant mass distribution for the $J/\psi \omega$ final state. The solid line represents the total fit function. The dashed line is the background contribution. The solid histogram is the non $J/\psi \omega$ background estimated from sidebands. The vertical dashed line is placed at the X(3872) nominal mass.}
\end{figure}

 \section{Search for the $Z_{1}(4050)^{+}$ and $Z_{2}(4250)^{+}$}
\begin{figure}[htb]
  \centering
 \includegraphics[height=.4\textheight]{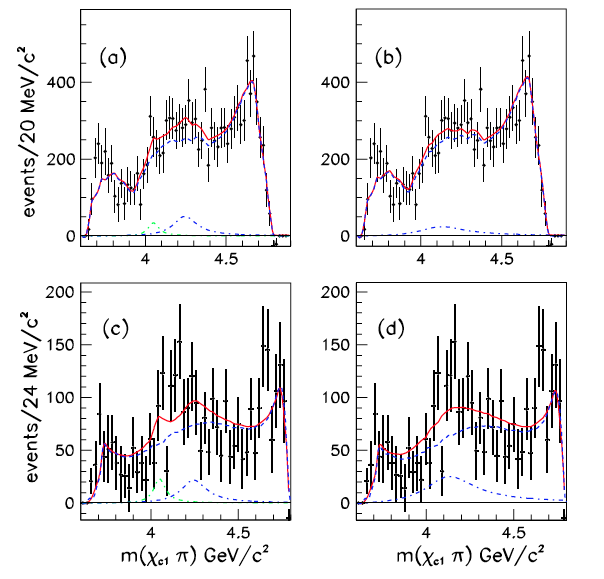}
  \label{fig4}
  \caption{(a),(b) Background-subtracted and efficiency-corrected $\chi_{c1}\pi$ mass distribution for $B\to \chi_{c1}K\pi$. (a) Fit with the $Z_{1}(4050)^{+}$ and $Z_{2}(4250)^{+}$ resonances. (b) Fit with only the $Z_{1}(4050)^{+}$ resonance. (c),(d) Efficiency-corrected and background-subtracted $\chi_{c1}\pi$ mass distribution in the $K\pi$ mass region where Belle found the maximum resonance activity: $1.0~<~m^{2}(K\pi)~<~1.75{\mathrm GeV^{2}/c^{4}}$. (c) Fit with $Z_{1}(4050)^{+}$ and $Z_{2}(4250)^{+}$ resonances. (d) Fit with only the $Z(4150)^{+}$ resonance. The dot-dashed curves indicate the fitted resonant contributions.}
  \end{figure}

In 2008 the Belle Collaboration reported the observation of a resonance-like structure called the $Z(4430)^{+}$ decaying to $\psi(2S)\pi^{+}$ studying of the process $B\to \psi(2S)K\pi$ \cite{zbelle}. This claim generated a great deal of interest \cite{maiani,karlip} since such a state must have a minimum quark content $c\bar{c}\bar{d}u$, and thus would represent an unequivocal manifestation of four-quark meson state. The \babar collaboration searched the  $Z(4430)^{+}$ in an analysis similar to that of Belle in the process  $B\to \psi(2S)K\pi$ and also in $B\to J/\psi K\pi$ \cite{zbabar}
but without finding any structure neither in the $\psi(2S) \pi$ nor in the $J/\psi \pi$. Recently Belle performed an amplitude analysis of the $J/\psi K\pi$ \cite{zcharm} system finding no significant evidence of the $Z(4430)^{+}$ in agreement with \babar. In 2009 the Belle Collaboration reported also the observation of two additional resonance-like structures similar to the $Z(4430)^{+}$ in the study of the $\bar{B}^{0}\to \chi_{c1}K^{-}\pi^{+}$ \cite{z12belle}. These new structures were labeled as the $Z_{1}(4050)^{+}$ and $Z_{2}(4250)^{+}$, both decaying to $\chi_{c1}\pi^{+}$.\\
 \babar, using a data sample of 429 $\mathrm{fb^{-1}}$, has recent searched for the $Z_{1}(4050)^{+}$ and $Z_{2}(4250)^{+}$in the process $\bar{B}^{0}\to \chi_{c1}K^{-}\pi^{+}$ and in the decay $B^{+}\to K_{s}^{0}\chi_{c1}\pi^{+}$ \cite{z12babar}, with the $\chi_{c1} \to J/\psi \gamma$. In the \babar analysis the $\chi_{c1}\pi^{+}$ mass distribution, after background subtraction and efficiency-correction, has been modeled using the angular information from the $K\pi$ mass distribution that has been represented using the Legendre polynomial moments. 
The excellent description of the $\chi_{c1}\pi^{+}$ mass distribution given by this analysis approach shows that there is no need for any additional 
resonance to model the distribution.  
Figure~\ref{fig4} shows the result of the fit done on the $\chi_{c1}\pi^{+}$ mass spectrum using two or one scalar Breit-Wigners with parameters fixed to the Belle measurement. In all the fit cases there are no significant resonant structures, since the statistical significance obtained is very low, less than $2 \sigma$. The Upper limits (ULs) at the 90 \% CL on the branching fractions are for the one resonance fit:
${\cal B}(\bar{B}^{0} \to Z^{+}K^{-}) \times {\cal B}(Z^{+} \to \chi_{c1}\pi^{+})<4.7 \times 10^{-5}$  while for the two resonances fit: 
${\cal B}(\bar{B}^{0} \to Z_{1}^{+}K^{-}) \times {\cal B}(Z^{+}_{1} \to \chi_{c1}\pi^{+})<1.8 \times 10^{-5}$ and ${\cal B}(\bar{B}^{0} \to Z^{+}_{2}K^{-}) \times {\cal B}(Z^{+}_{2} \to \chi_{c1}\pi^{+})<4.0 \times 10^{-5}$.

  \section{Study of the $J/\psi\pi^{+}\pi^{-}$ system via Initial State Radiation (ISR)}
\begin{figure}[htb]
  \centering
 \includegraphics[height=.2\textheight]{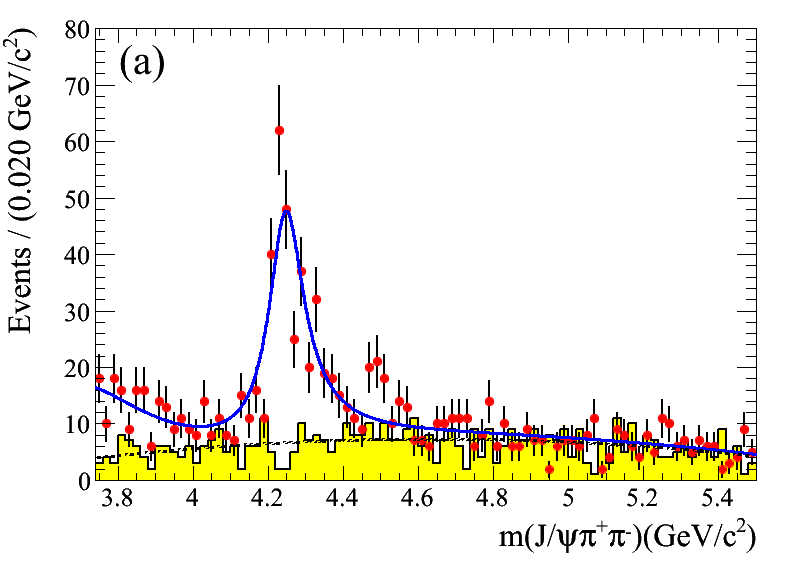}
 \includegraphics[height=.2\textheight]{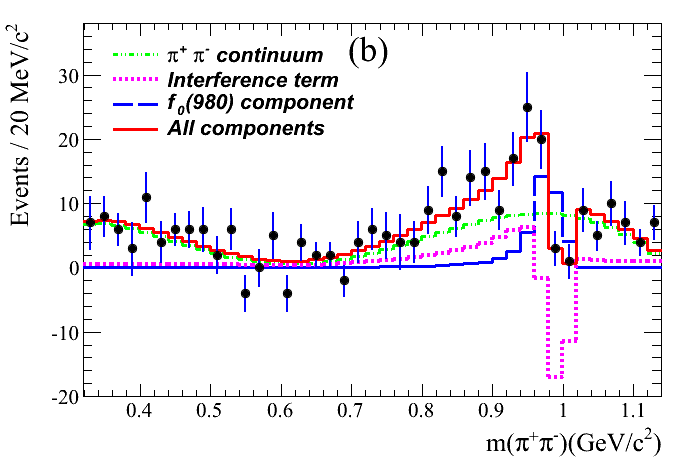}
  \label{fig5}
  \caption{(a): the $J/\psi \pi^{+}\pi^{-}$ mass spectrum from 3.74 ${\mathrm{GeV/c^{2}}}$ to 5.5 ${\mathrm{GeV/c^{2}}}$; the points represent the data and the shaded histogram is the background from the $J/\psi$~sideband; the solid curve represent the fit result. (b) the $\pi^{+}\pi^{-}$ distribution from the Y(4260) decay  to $J/\psi \pi^{+}\pi^{-}$. The solid curve represent the fit using the model described in the text.}
    \end{figure}

In 2005 \babar discovered the Y(4260) in the process $e^{+}e^{-} \to \gamma_{ISR} Y(4260)$, with the $Y(4260) \to J/\psi\pi^{+}\pi^{-}$ \cite{Ybabar}.
Since it is produced directly in $e^{+}e^{-}$ annihilation it has $J^{PC}=1^{--}$.
The observation of the decay
mode $J/\psi\pi^0\pi^0$ \cite{Ypi0} established that it has zero isospin. 
However it is not observed to decay to $D^*\bar{D^*}$ \cite{Ydd}, nor to $D_s^*\bar{D^*_s}$ \cite{Yds}, so that its properties do not lend 
itself to a simple charmonium interpretation, and its nature is still unclear.
A subsequent Belle analysis \cite{Ybelle} of the same final state suggested also the existence
of an additional resonance around 4.1 GeV/c$^2$ that they dubbed the Y(4008).
\babar has performed recently a new analysis \cite{Ybabarnew}
 of this process using 454 ${\mathrm fb^{-1}}$. In this new study the region below 4.0 ${\mathrm GeV/c^{2}}$ has been studied for the first time. As shown on Figure \ref{fig5}(a) in that region there is an excess of events above the $J/\psi$ sidebands background. To understand the nature of this contribution a detailed study of the $\psi(2S)$ line shape has been performed and the result shows that it is not possible to discount the presence of a $J/\psi\pi^{+}\pi^{-}$  continuum cross section in this region. Figure \ref{fig5}(a) shows the fit to the $J/\psi\pi^{+}\pi^{-}$ mass distribution. A clear signal for the Y(4260) is seen; the values obtained from an unbinned-maximum-likelihood fit are: $m_{Y(4260)}=4244 \pm 5 \pm4~{\mathrm MeV/c^{2}}$, $\Gamma_{Y(4260)}=114^{+16}_{-15} \pm 7$ MeV and $\Gamma_{ee} \times {\cal B}(J/\psi \pi^{+}\pi^{-}) =9.2 \pm 0.8 \pm 0.7$~eV. There is no evidence for the Y(4008) seen by Belle \cite{Ybelle}. \\ In the new \babar analysis a detailed study of the $\pi^{+}\pi^{-}$ system from the Y(4260) decay  to $J/\psi \pi^{+}\pi^{-}$ has been performed. The $\pi^{+}\pi^{-}$ mass distribution shown in Figure~\ref{fig5}(b) seems to peak around the $f_{0}(980)$ mass; however the peak is displaced from the nominal $f_{0}(980)$ position, since it is around 940 ${\mathrm{MeV/c^{2}}}$. The fact that the peak is displaced and the particular shape of $m(\pi^{+}\pi^{-})$ distribution may suggest a possible interference between the $f_{0}(980)$ and $m(\pi^{+}\pi^{-})$ continuum. To test this possibility the $f_{0}(980)$ line shape is taken from the \babar analysis \cite{antimo} of the $D_{S}^{+}\to \pi^{+}\pi^{-}\pi^{+}$ and this amplitude has been used in a simple model to describe the $\pi^{+}\pi^{-}$ mass distribution: $|\sqrt{pol} +e^{i\phi}F_{f_{0}(980)}|^{2}$ where $pol$ is a polynomial function used to describe the  $m(\pi^{+}\pi^{-})$ continuum and 
$F_{f_{0}(980)}$ is the amplitude from the $D_{S}^{+}\to \pi^{+}\pi^{-}\pi^{+}$ \cite{antimo} analysis; $\phi$ allows for a phase difference between these amplitudes. The result of this study is shown in Figure~\ref{fig5}(b) and indicates that if there is a real $f_{0}(980)$ contribution to the decay of the Y(4260) to~$J/\psi\pi^{+}\pi^{-}$ its contribution is not dominant: $\frac{{\cal B}(Y_{4260} \to J/\psi f_{0}(980), ~f_{0}(980)\to \pi^{+}\pi^{-})}{{\cal B}(Y_{4260}   \to J/\psi \pi^{+}\pi^{-})}=(17 \pm 13) \%$.

  \begin{figure}[htb]
  \centering
 \includegraphics[height=.22\textheight]{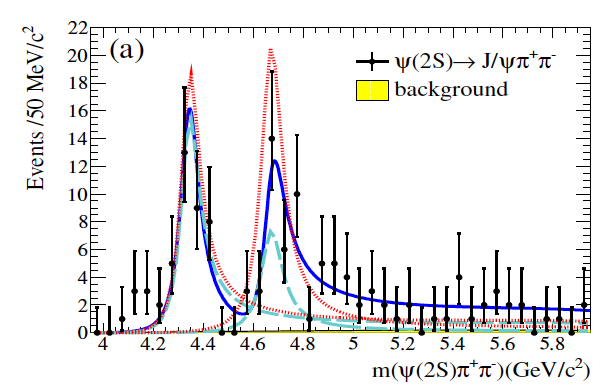}
  \label{fig6}
  \caption{The $\psi(2S)\pi^{+}\pi^{-}$ invariant mass distribution from 3.95 to 5.95 ${\mathrm GeV/c^{2}}$ for the $\psi(2S) \to J/\psi \pi^{+}\pi^{-}$; the points with error bars represent the data in the $\psi(2S)$ signal region, and the shaded histogram is the background estimated from the $\psi(2S)$ sideband regions. The solid curve show the result of the fit. }
  \end{figure}

\section{Study of the $\psi(2S)\pi^{+}\pi^{-}$ system via Initial State Radiation (ISR)}
In addition to the Y(4260), two more $J^{PC}=1^{--}$ states, the Y(4360) and the Y(4660) have been reported in ISR production $e^{+}e^{-}\to \psi(2S) \pi^{+}\pi^{-}$  \cite{Y2babar,Y2belle}. While the Y(4360) was discovered by \babar~\cite{Y2babar} and then confirmed by Belle~\cite{Y2belle} the Y(4660) was only observed by the Belle Collaboration. \babar performed a new analysis~\cite{Y4660BaBar} using all its available dataset collected at the $\Upsilon (nS)$,~n=2,3,4; that corresponds to an integrated luminosity of 520${\mathrm{fb^{-1}}}$. The $\psi(2S)\pi^{+}\pi^{-}$ mass spectrum for the $\psi(2S) \to J/\psi \pi^{+}\pi^{-}$ is reported in Figure~\ref{fig6}. \babar observes two resonant structures, that have been interpreted as the Y(4360) and the Y(4660), respectively. The parameters values obtained from an unbinned-maximum-likelihood are for the first resonance $m_{Y(4360)}=4340 \pm 16 \pm 9~{\mathrm MeV/c^{2}}$, $\Gamma_{Y(4360)}=94 \pm 32  \pm 13$~MeV, and for the second one $m_{Y(4660)}=4669 \pm 21 \pm 3~{\mathrm MeV/c^{2}}$, $\Gamma_{Y(4660)}=104 \pm 48 \pm 10$~MeV.

\end{document}